# Buffered Aloha with *K*-Exponential Backoff Part II: Delay Analysis

Lin Dai, *Member, IEEE*, and Tony T. Lee, *Fellow, IEEE*

*Abstract*— This paper presents the delay analysis for buffered Aloha networks with *K*-Exponential Backoff. Mean access delay and mean queueing delay are derived and demonstrated via the examples of Geometric Retransmission (*K*=1) and Exponential Backoff (*K*=∞). The comparison shows that higher delay is incurred with Geometric Retransmission when the aggregate input rate is small, and the delay gap is enlarged as the number of nodes *n* increases. With a high traffic input rate, however, the delay performance with Exponential Backoff severely deteriorates. The mean queueing delay will be unbounded if the aggregate input rate exceeds 0.3.

We also extend the analysis to the contention-window-based backoff model which is widely adopted in practical MAC protocols. It will be revealed that both the retransmission-probability-based and the contention-window-based models exhibit the same stable region and achieve similar queueing performance in most cases, which justifies the intuition that was taken but remained unverified in previous studies: the retransmission-probability-based backoff model can serve as a good approximation of the contention-window-based one.

*Index Terms*—Slotted Aloha, exponential backoff, geometric retransmission, stability, queueing delay, access delay, contention window

## I. INTRODUCTION

This is Part II of the paper series that studies the performance of buffered Aloha networks with *K*-Exponential Backoff. Stability and throughput analysis has been presented in Part I. Delay performance will be further investigated in this paper.

Delay analysis of buffered Aloha networks has long remained an open problem. In a buffered Aloha network, each node is equipped with an infinite or finite queue, and shares a common server. The total queueing delay of each packet consists of two parts: 1) waiting time, which is the time interval from the packet's arrival to the instant that it becomes the head-of-line (HOL) packet; and 2) access delay, which is the time interval from the instant that it becomes the HOL packet to its successful transmission. Note that due to contention, the server may be paralyzed if the attempt rate is too high. In that case, the whole network will become unstable with unbounded delay.

Intuitively, a buffered Aloha system can be modeled as a multi-dimensional Markov chain with state vector ($C_1$, $C_2$, …, $C_n$), where $C_i$ represents the queue length of node *i*, *i*=1,…, *n* [1]. The ultimate solution for the case of *n*=2 has been characterized in [2]; nevertheless, the approach becomes intractable as the dimension exceeds three [3-6]. To simplify the analysis, various approximate methods have been developed to tackle the coupled queues [7-18]. Although most of them were customized for specific system configurations, the idea of *decomposition* gradually becomes a consensus: each node can be approximately treated as an independent queueing system with identically distributed service time [13-18]. The key to delay analysis is the characterization of service time distribution, which unfortunately remains quite elusive.

In fact, stability analysis is a prerequisite for delay analysis in random access networks. In contrast to traditional queueing systems, the service rate of each single queue in a random access network is determined by the aggregate activities of all the nodes. The queueing delay will become infinite if the system collapses, i.e., the aggregate service rate drops below the aggregate input rate. It is therefore important to characterize the system stable region first.

The stability analysis for *n*-node buffered Aloha networks with *K*-Exponential Backoff has been presented in Part I of the paper series, where the original Slotted Aloha (which is referred to as Geometric Retransmission) and Exponential Backoff are included as two special examples with the cutoff phase *K*=1 and ∞, respectively. Assume that each node is equipped with an infinite buffer and has Bernoulli arrivals with an input rate of $\lambda$ packets per time slot. We characterize the stable region *S* of retransmission factor *q*, within which a network throughput of $\hat{\lambda}_{out} = \hat{\lambda} = n\lambda$ can be achieved for any aggregate input rate $0 < \hat{\lambda} \leq \hat{\lambda}_{max\_S} = e^{-1}$.

It will be further revealed in this paper that although the throughput stability can be always achieved with any $q \in S$, the delay performance is critically dependent on the value of retransmission factor *q*, and might be drastically different with various backoff schemes. For instance, with Geometric Retransmission (*K*=1), the mean queueing delay of each packet will quickly decrease as the retransmission probability increases, and is minimized when *q* reaches the upper bound of the stable region $S^{Geo}$. The maximum stable throughput $\hat{\lambda}_{max\_S} = e^{-1}$ can be achieved with *q*=1/*n*, and the corresponding queueing delay linearly increases with the number of nodes *n*.

In contrast, the delay performance of Exponential Backoff ($K=\infty$) networks is insensitive to the network population $n$ when the aggregate input rate $\hat{\lambda}$ is small. Much lower queueing delay is therefore incurred compared to Geometric Retransmission, especially when the number of nodes $n$ is large. Nevertheless, the delay performance deteriorates sharply as the traffic level increases. Our analysis shows that with $q \in S^{Exp}$, the Exponential Backoff system will become quasi-stable if the aggregate input rate $\hat{\lambda} \geq 0.3$. The maximum stable throughput $\hat{\lambda}_{\max\_S}=e^{-1}$ is achieved at the cost of infinite mean queueing delay.

Note that in spite of severely penalized delay performance, Exponential Backoff networks provide the best robustness: a stable throughput can be achieved with the retransmission factor $q$ fixed to be $1-e^{-1}$ regardless of the traffic input rate or the network population. With Geometric Retransmission, however, the stable region rapidly shrinks as the number of nodes $n$ increases. Any slight change of network population $n$ will lead to a risk of system collapse if the retransmission factor $q$ is not updated accordingly.

The performance comparison of Geometric Retransmission ($K=1$) and Exponential Backoff ($K=\infty$) suggests that for general $K$-Exponential Backoff systems, the cutoff phase $K$ can serve as a leverage to strike a balance between queueing performance and system robustness. Given the retransmission factor $q$, increasing $K$ can greatly improve the maximum stable throughput that a network can achieve. On the other hand, it may also impair the queueing delay performance when the aggregate input rate $\hat{\lambda}$ is high. The optimal value of cutoff phase $K$ is subject to the system requirements such as the range of traffic input rate and the delay constraint.

Note that in practical MAC protocols, a contention-window mechanism is usually adopted [8, 17-20], which is slightly different from the retransmission-probability-model supposed in most analytical studies [2-7, 21-24]. Our analysis will further demonstrate that both models indeed share the same stable region. Lower queueing delay is incurred by the window-based backoff model; nevertheless, the delay gap is quite small unless the cutoff phase $K$ and the traffic input rate $\hat{\lambda}$ are both high.

The remainder of this paper is organized as follows. Section II presents the delay analysis of *K*-Exponential Backoff, where Geometric Retransmission and Exponential Backoff are investigated as exemplary cases, and the tradeoff between queueing delay performance and system robustness is characterized. The analysis is further extended to the contention-window-based backoff model in Section III. Conclusions are summarized in Section IV.

## II. QUEUEING DELAY OF *K*-EXPONENTIAL BACKOFF

As demonstrated in Part I, an *n*-node buffered Aloha network can be modeled as a non-work-conservative *n*-queue-single-server system. Each node is supposed to be equipped with an infinite buffer and the coupled queues are decomposed into independent FIFO queues with Bernoulli arrivals of rate $\lambda$ packets per time slot. These decomposed Geo/G/1 queues are then hinged together by the probability of success $p$ of the HOL packets. Fig. 1 presents the phase model that was established in Part I to describe the state transition process of each individual HOL packet. With *K*-Exponential Backoff, the transmission probability of a HOL packet in phase $i$ is $q^i$, $i=0,1,\ldots,K$.

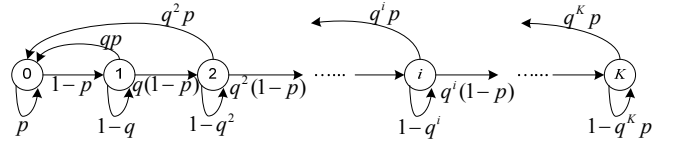

Fig. 1. State transition diagram of HOL packets.

It can be seen from Fig. 1 that the service time $X$ of a HOL packet is the sum of the time spent in successive phases, which is determined by both the probability of success $p$ and the retransmission factor $q$. Let $X_i$ denote the time spent from the beginning of phase $i$ until the service completion, and $Y_i$ be the time spent in phase $i$. We have

$$X_i = \begin{cases} Y_i & \text{with probability } p \\ Y_i + X_{i+1} & \text{with probability } 1-p \end{cases}, \quad (1)$$

$i=0,1,\ldots,K-1$, and $X_K=Y_K$.

With *K*-Exponential Backoff, the sojourn time at phase $i$, $Y_i$, is a geometric distributed random variable with the probability generating function

$$\begin{cases} G_{Y_i}(z) = \dfrac{q^i z}{1-(1-q^i)z}, \quad i=0,1,\ldots,K-1 \\ G_{Y_K}(z) = \dfrac{pq^K z}{1-(1-pq^K)z}. \end{cases} \quad (2)$$

The probability generating function of $X_i$ can be then given as follows:

$$\begin{cases} G_{X_i}(z) = \dfrac{q^i z}{1-(1-q^i)z}\left[p+(1-p)G_{X_{i+1}}(z)\right], \quad i=0,1,\ldots,K-1 \\ G_{X_K}(z) = \dfrac{pq^K z}{1-(1-pq^K)z}. \end{cases} \quad (3)$$

From (3), we immediately have

$$G'_{X_0}(1) = 1 + \frac{1-p}{q-(1-p)} - \left(\frac{1-p}{q-(1-p)} - \frac{1-p}{p}\right)\cdot\left(\frac{1-p}{q}\right)^K \quad (4)$$

and

$$G''_{X_0}(1) = \frac{2(1-p)q}{[q-(1-p)]\cdot[q^2-(1-p)]} +$$
$$\left(\frac{1-p}{q^2}\right)^K \cdot \left\{\frac{2q}{q-(1-p)}\left(q^K - \frac{q^2}{q^2-(1-p)}\right) + \right.$$
$$\left. \frac{2(1-q^K p)}{p^2} - \frac{q^K-1}{1-q^{-1}}\left(\frac{1-p}{q-(1-p)} - \frac{1-p}{p}\right)\right\}. \quad (5)$$

The mean and variance of the service time $X$ are then given by
$$\mathrm{E}[X] = G'_{X_0}(1), \quad (6)$$
and
$$\mathrm{var}[X] = G''_{X_0}(1) + G'_{X_0}(1) - G'_{X_0}(1)^2, \quad (7)$$
respectively.

Finally, the mean queueing delay of input packets can be obtained by substituting (4-5) into the Pollaczek-Kintchine (P-K) formula given in Appendix I:

$$\mathrm{E}[T] = G'_{X_0}(1) + \frac{\lambda G''_{X_0}(1)}{2(1-\lambda G'_{X_0}(1))}. \quad (8)$$

It can be seen from (6) and (8) that mean access delay E[X] and mean queueing delay E[T] are both determined by the probability of success $p$ and the retransmission factor $q$. The stable region $S$ of retransmission factor $q$ has been characterized in Part I, within which a stable throughput can be achieved. Apparently, the mean queueing delay becomes infinite if the retransmission factor $q$ falls outside the stable region $S$.

Define $S_D$ as the delay-stable region of retransmission factor $q$, within which the mean queueing delay of input packets is finite:

$$S_D = \{q \mid \mathrm{E}[T] < \infty\}. \quad (9)$$

It is plain to see that $S_D$ is a subset of the stable region $S$, i.e., $S_D \subseteq S$, because a stable throughput is a necessary but not sufficient condition for finite mean queueing delay E[T]<∞. We are only interested in the delay performance within the delay-stable region $S_D$. We will take the examples of Geometric Retransmission (K=1) and Exponential Backoff (K=∞) to demonstrate the above results.

### A. Queueing Delay of Geometric Retransmission

When the cutoff phase K=1, the first and second moments of service time X can be obtained from (4-5) as

$$G'_{X_0}(1) = 1 + \frac{1-p}{pq} \quad (10)$$

and

$$G''_{X_0}(1) = \frac{2(1-p)}{(pq)^2}. \quad (11)$$

The mean access and mean queueing delay of Geometric Retransmission are then given by

$$\mathrm{E}[X]^{Geo} = 1 + \frac{1-p}{pq}, \quad (12)$$

$$\mathrm{E}[T]^{Geo} = 1 + \frac{1}{pq - \frac{\lambda(1-p)}{1-\lambda}} - \frac{1}{q}. \quad (13)$$

It has been demonstrated in Part I that an $n$-node network with Geometric Retransmission has a stable throughput if and only if the retransmission factor $q$ is selected from the stable region

$$S^{Geo} = \left[\frac{\hat{\lambda}(1-p_L)}{p_L(n-\hat{\lambda})}, \frac{-\ln p_S}{n}\right], \quad (14)$$

and the probability of success $p$ will converge to the desired stable point $p_L$. Delay stability can be achieved within the stable region $S^{Geo}$ except when $q$ reaches the lower bound, with which the offered load of each queue $\rho$=1. As a result, the delay-stable region of Geometric Retransmission is given by

$$S_D^{Geo} = \left(\frac{\hat{\lambda}(1-p_L)}{p_L(n-\hat{\lambda})}, \frac{-\ln p_S}{n}\right], \quad (15)$$

and the corresponding maximum stable throughput that the network can achieve with finite mean queueing delay, $\hat{\lambda}_{max\_D}^{Geo}$, is given by

$$\hat{\lambda}_{max\_D}^{Geo} = \hat{\lambda}_{max\_S}^{Geo} = e^{-1}. \quad (16)$$

The desired stable point $p_L$ is solely determined by the aggregate input rate $\hat{\lambda} = n\lambda$. According to (12-13), both mean access delay E[X] and mean queueing delay E[T] will decrease as the retransmission factor $q$ increases inside the delay-stable region $S_D^{Geo}$. The minimum mean access delay and the minimum mean queueing delay of Geometric Retransmission can be then obtained as

$$\min_{q \in S_D^{Geo}} \mathrm{E}[X]^{Geo} = 1 + \frac{1-p_L}{p_L} \cdot \frac{n}{-\ln p_S}, \quad (17)$$

$$\min_{q \in S_D^{Geo}} \mathrm{E}[T]^{Geo} = 1 + n\left(\frac{1}{-p_L \ln p_S - \frac{\hat{\lambda}(1-p_L)}{1-\hat{\lambda}/n}} - \frac{1}{-\ln p_S}\right), \quad (18)$$

which can be achieved when

$$q = -\frac{1}{n}\ln p_S = -\frac{1}{n}W_{-1}(-\hat{\lambda}). \quad (19)$$

It is clear from (17-18) that both the minimum mean access delay and the minimum mean queueing delay increase *linearly* with the number of nodes $n$. They also increase with the aggregate input rate $\hat{\lambda}$.

Fig. 2 shows the minimum mean access delay and minimum mean queueing delay versus $\hat{\lambda}$ under different values of $n$. It can be seen from Fig. 2 that the queueing delay is close to the access delay only when traffic is light. A sharp increase of delay (both access delay and queueing delay) can be observed when $\hat{\lambda}$ approaches the maximum $\hat{\lambda}_{max\_D}^{Geo} = e^{-1}$, and the delay is doubled when the number of nodes $n$ increases from 50 to 100. The simulation results presented in Fig. 2 perfectly agree with the analysis.

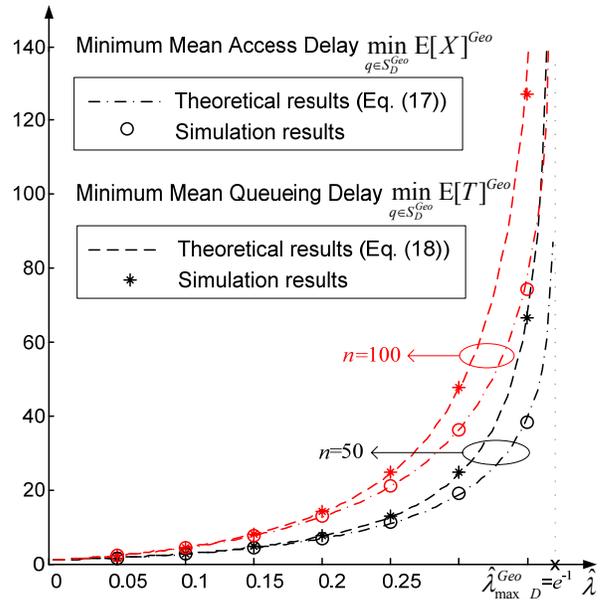

Fig. 2. Minimum mean queueing delay and minimum mean access delay versus aggregate input rate $\hat{\lambda}$ with Geometric Retransmission achieved at $q=-\ln p_S/n$.

To achieve the best delay performance, however, the retransmission factor $q=-\frac{1}{n}W_{-1}(-\hat{\lambda})$ needs to be adjusted whenever the traffic level changes. Even a slight increment of $\hat{\lambda}$ could lead to system breakdown if $q$ is not updated in time. In practice, it is desired that the network can operate in stable states under any traffic input rate $0<\hat{\lambda}\leq\hat{\lambda}_{max\_S}^{Geo}=e^{-1}$. According to (14), this can be achieved with the retransmission factor $q$ set to be $q_m^{Geo}=1/n$.

With $q=q_m^{Geo}=1/n$, the corresponding mean access delay and mean queueing delay are given by:

$$E[X]_{q=q_m}^{Geo}=1+n\cdot\frac{1-p_L}{p_L}, \quad (20)$$

$$E[T]_{q=q_m}^{Geo}=1+n\left(\frac{1}{p_L-\frac{\hat{\lambda}(1-p_L)}{1-\hat{\lambda}/n}}-1\right). \quad (21)$$

Both of them again increase with the number of nodes $n$ and the aggregate input rate $\hat{\lambda}$.

The minimum mean queueing delay and the mean queueing delay with $q=q_m^{Geo}$ are presented in Fig. 3. It can be clearly seen that the delay gap increases with the aggregate input rate $\hat{\lambda}$. A closer look at (18) and (21) further suggests that the gap also increases with the number of nodes $n$. As we have illustrated, $q_m^{Geo}=1/n$ is a more practical choice. Part of the delay performance has to be sacrificed in return of better system robustness.

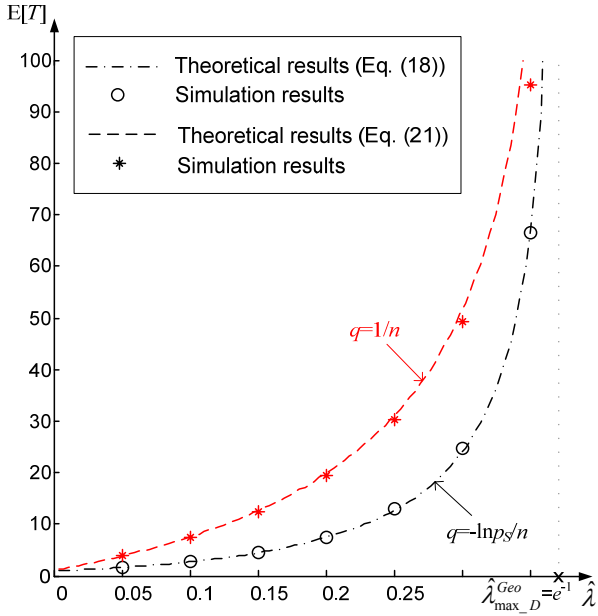

Fig. 3. Mean queueing delay with Geometric Retransmission under different values of retransmission factor $q$. $n=50$.

### B. Queueing Delay of Exponential Backoff

When the cutoff phase $K=\infty$, the first and second moments of service time $X$ can be obtained from (4-5) as

$$G'_{X_0}(1)=1+\frac{1-p}{q-(1-p)} \quad (22)$$

and

$$G''_{X_0}(1)=\frac{2(1-p)q}{[q-(1-p)]\cdot[q^2-(1-p)]}, \quad (23)$$

if the retransmission factor satisfies

$$q>\sqrt{1-p}. \quad (24)$$

Otherwise the second moment of $X$ will become infinite. The corresponding mean access and mean queueing delay of Exponential Backoff are given by

$$E[X]^{Exp}=1+(1-p)/(p+q-1), \quad (25)$$

$$E[T]^{Exp}=1+\frac{1-p}{p+q-1}+\frac{\lambda(1-p)q}{(p+q-1-\lambda q)(p+q^2-1)}. \quad (26)$$

#### B.1 Delay-Stable Region of Exponential Backoff

The stable region of Exponential Backoff has been characterized in Part I. In particular, with a retransmission factor $q$ selected from the absolute-stable region $S_L^{Exp}$, the network is guaranteed to operate at the desired stable point $p_L$. When the network shifts to the undesired stable point $p_A\approx1$-$q$, a stable throughput can still be achieved if the retransmission factor $q$ is selected from the stable region $S_A^{Exp}$. The complete stable region is therefore given by

$$S^{Exp}=S_L^{Exp}\bigcup S_A^{Exp}=[1-p_L,1-p_S]. \quad (27)$$

If stable throughput is the only concern, it does not matter whether the probability of success $p$ converges to $p_L$ or $p_A$: throughput stability can always be achieved as long as the retransmission factor $q$ is selected from the stable region $S^{Exp}$. Nevertheless, the delay performance is drastically different at the above two stable points. It is shown in Appendix II that when the probability of success $p$ converges to the undesired stable point $p_A$, (24) cannot be satisfied, implying that the mean queueing delay will become unbounded.

Similar to Geometric Retransmission, the Exponential Backoff system is guaranteed to operate at the desired stable point $p_L$ if the retransmission factor $q$ is selected from the absolute-stable region $S_L^{Exp}$, at which delay stability is achievable. Unfortunately, the absolute-stable region of Exponential Backoff will rapidly shrink as the number of nodes $n$ increases. With $n=100$, for instance, $S_L^{Exp}$ becomes an empty set when the aggregate input rate $\hat{\lambda}$ exceeds 0.045.

To characterize the delay-stable region of Exponential Backoff with a large number of nodes $n$, we need to determine how the probability of success $p$ varies with the retransmission factor $q$ outside the absolute-stable region, which is difficult to be obtained. Nevertheless, it is shown in Appendix III that with

$$\frac{-\ln p_S}{1-\ln p_S}\leq q\leq 1-p_S, \quad (28)$$

the Exponential Backoff system will operate at the undesired stable point $p_A$ with high probability, at which the mean queueing delay is unbounded. Moreover, if the system operates at the desired stable point $p_L$, we can see from (24) that with

$$1-p_L\leq q\leq\sqrt{1-p_L}, \quad (29)$$

the mean queueing delay will also become infinite.

As shown in Fig. 4, despite a stable throughput, the mean queueing delay will grow unboundedly if the retransmission factor $q$ falls into the shadowing area, indicating that the network is quasi-stable.

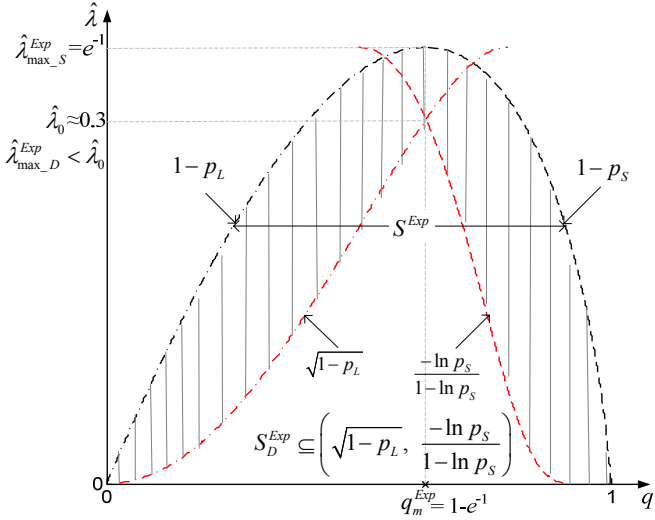

Fig. 4. Stable region of Exponential Backoff.

By combining (28) and (29), we can conclude that for large number of nodes $n$, the delay-stable region of Exponential Backoff should satisfy:

$$S_D^{Exp} \subseteq \left( \sqrt{1-p_L}, \frac{-\ln p_S}{1-\ln p_S} \right). \quad (30)$$

It can be readily seen from (30) and Fig. 4 that for Exponential Backoff, the maximum stable throughput with finite mean queueing delay, $\hat{\lambda}_{max\_D}^{Exp}$, satisfies

$$\hat{\lambda}_{max\_D}^{Exp} < \hat{\lambda}_0 < \hat{\lambda}_{max\_S}^{Exp} = e^{-1}, \quad (31)$$

where $\hat{\lambda}_0$ is the root of the following equation

$$\sqrt{1-p_L} = \frac{-\ln p_S}{1-\ln p_S}. \quad (32)$$

Infinite mean queueing delay will be incurred with Exponential Backoff when the aggregate input rate $\hat{\lambda}$ exceeds $\hat{\lambda}_0 \approx 0.3$.

It has been shown in Fig. 17 of Part I that with $\hat{\lambda} = 0.3$, the Exponential Backoff system will operate at the undesired stable point $p_A$ with any retransmission factor $q \in S^{Exp}$. As we have mentioned in Section IV. B, Part I, "capture phenomenon" [25-27] occurs when the Exponential backoff system operates at the undesired stable point $p_A$. In that case, many nodes will be pushed to large phases with extremely small retransmission probabilities such that the node who once succeeds can dominate the channel for a long time and produce a continuous stream of packets. The network throughput can still strike a balance, but the output process is no longer stationary. The queue length of each single node will vary violently and fail to converge as time grows.

Fig. 5 plots the mean queueing delay of Exponential Backoff with the traffic input rate $\hat{\lambda} = 0.3$ and the retransmission factor $q = 1 - e^{-1}$. It can be clearly seen that the mean queueing delay constantly varies with time. It rapidly mounts up once some HOL packets are blocked in deep phases, and starts declining when any of those nodes finally captures the channel and releases its accumulated packets. As time elapses, the mean queueing delay fails to follow any bound, because the retransmission probability of each HOL packet can be arbitrarily small. The second moment of service time in this case has indeed diverged. As a result, the mean queueing delay grows unboundedly, although a stable throughput can still be achieved in the long run.

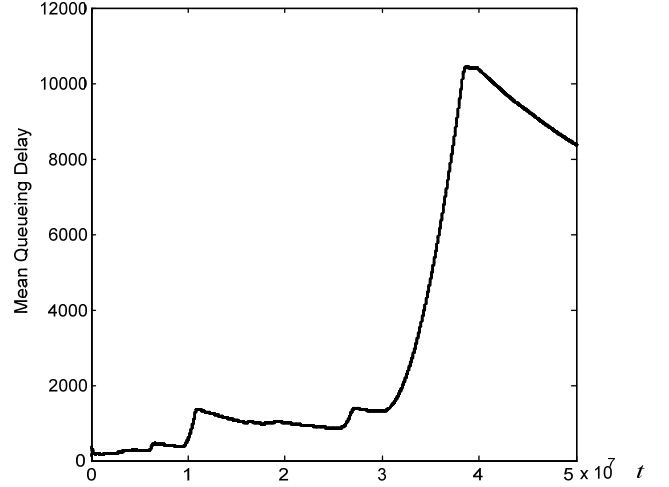

Fig. 5. Mean queueing delay of Exponential Backoff with $n=100$, $q=1-e^{-1}$ and $\hat{\lambda}=0.3$.

### B.2 Delay Performance with $q=q_m^{Exp}=1-e^{-1}$

In contrast to Geometric Retransmission, the optimal delay performance of Exponential Backoff cannot be obtained as the delay-stable region with a large number of nodes $n$ is unknown. In the following, we will focus on the delay performance when $q=q_m^{Exp}=1-e^{-1}$, with which a stable throughput can be achieved under any traffic input rate $0 < \hat{\lambda} \le \hat{\lambda}_{max\_S}^{Exp} = e^{-1}$.

Suppose that the system operates at the desired stable point, i.e., $p=p_L$, with $q=q_m^{Exp}$ and $\hat{\lambda} < \hat{\lambda}_0$. The mean access and queueing delay can be obtained from (25-26) as

$$\mathrm{E}[X]_{q=q_m}^{Exp} = \frac{1-e^{-1}}{p_L - e^{-1}}, \quad (33)$$

$$\mathrm{E}[T]_{q=q_m}^{Exp} = \frac{1-e^{-1}}{p_L - e^{-1}} + \frac{\hat{\lambda}(1-p_L)(1-e^{-1})}{(n(p_L - e^{-1}) - \hat{\lambda}(1-e^{-1}))(p_L - 2e^{-1} + e^{-2})}. \quad (34)$$

It is interesting to see from (34) that with a large number of nodes $n$,

$$\mathrm{E}[T]_{q=q_m}^{Exp} \approx \mathrm{E}[X]_{q=q_m}^{Exp} = \frac{1-e^{-1}}{p_L - e^{-1}}, \quad (35)$$

implying that the delay does not vary with the number of nodes $n$ if the system operates at the desired stable point $p_L$. This can be clearly observed in Fig. 6.

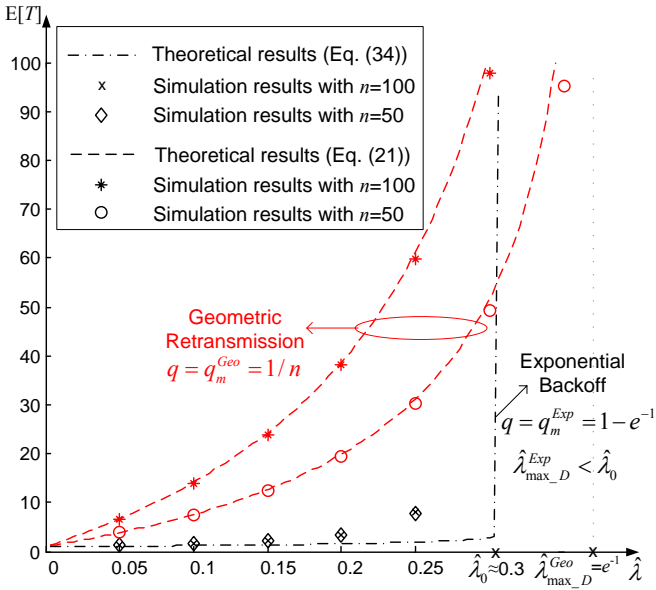

Fig. 6. Mean queueing delay versus $\hat{\lambda}$ with Exponential Backoff and Geometric Retransmission when $q=q_m$.

Note that the gap between simulation and analytical results becomes noticeable as the aggregate input rate $\hat{\lambda}$ approaches $\hat{\lambda}_0$. A closer observation indicates that the correlation among queues will become significantly high as the traffic input rate $\hat{\lambda}$ increases. The mounting correlation will lead to a departure of the probability of success from the theoretical value of $p_L$, which is derived based on the assumption of independent queues. When the aggregate input rate $\hat{\lambda}$ exceeds $\hat{\lambda}_0$, the probability of success will converge to the undesired stable point $p_A$, and the mean queueing delay will grow unboundedly.

It can be seen from Fig. 6 that the delay performance of Exponential Backoff is highly polarized. When the traffic input rate $\hat{\lambda}$ is lower than $\hat{\lambda}_0$, significant improvement can be observed compared to Geometric Retransmission. The delay gap becomes even larger as the number of nodes *n* increases, because the mean queueing delay with Geometric Retransmission linearly increases with *n*. Nevertheless, Geometric Retransmission has much better queueing performance when the traffic level is high. Finite mean queueing delay is incurred over the whole traffic range $0<\hat{\lambda}\leq\hat{\lambda}_{\max\_D}=e^{-1}$.

Although the Exponential Backoff system becomes quasi-stable when traffic is heavy, it provides the best robustness. As shown in Fig. 4, the stable region of Exponential Backoff does not vary with the number of nodes *n*. The retransmission factor *q* can be fixed to be $q_m^{Exp}=1-e^{-1}$ no matter how many nodes the network has. In contrast, the stable region of Geometric Retransmission will quickly shrink as the network population *n* increases. According to (14), the retransmission factor *q* has to be adjusted in the order of 1/*n* to achieve stability. This indicates that any increase of number of nodes *n* may lead to system collapse if *q* is not updated in time. Realtime feedback on the total number of nodes is critical for stabilizing the Geometric Retransmission systems.

### C. Tradeoff between Queueing Performance and Robustness

From the comparison on stability and delay performance of Geometric Retransmission (*K*=1) and Exponential Backoff (*K*=∞), we can see that with a small cutoff phase *K*, the system has good queueing performance but is prone to be unstable. Intuitively, nodes have to back off to larger phases to alleviate contention. If the cutoff phase *K* is small, all the nodes will quickly reach the maximum phase *K* and the retransmission probabilities cannot be reduced any more. As a result, the network has very limited capability to absorb the growing contention and will quickly collapse as the number of nodes *n* or traffic input rate $\hat{\lambda}$ increases. The positive side, however, is that the queueing performance is quite good because the retransmission probability of each HOL packet is lower bounded by $q^K$.

With *K* increasing, nodes will have more space to lower down their transmission requests as contention grows. An infinite cutoff phase *K* indicates that nodes can always back off to deeper phases to make the attempt rate arbitrarily small until the network is stabilized. That is why the stable region of Exponential Backoff does not vary with the number of nodes *n*. Nevertheless, some HOL packets may suffer from extremely long delay due to the small retransmission probabilities, and the queueing delay may become unbounded as the variance of service time diverges when the traffic input rate is too high.

For *K*-Exponential Backoff with 1<*K*<∞, the cutoff phase *K* can serve as a leverage to achieve a tradeoff between queueing performance and system robustness. For illustration, let us fix the retransmission factor *q* at $q_m^{Exp}=1-e^{-1}$. As we have shown in Section II. B, with Exponential Backoff, the maximum stable throughput that a network can achieve with $q=1-e^{-1}$ is

$$\hat{\lambda}_{\max\_S\_q_m^{Exp}}^{Exp} = e^{-1}. \qquad (36)$$

With Geometric Retransmission, however, the corresponding maximum stable throughput is given by

$$\hat{\lambda}_{\max\_S\_q_m^{Exp}}^{Geo} = n(1-e^{-1})\exp\{n(1-e^{-1})\} \qquad (37)$$

according to Eqs. (50) and (52) in Part I, which is close to zero when the number of nodes *n* is large. In other words, the Geometric Retransmission system will collapse almost under any non-zero traffic input rate $\hat{\lambda}>0$ if $q=q_m^{Exp}=1-e^{-1}$. It is clear that the Exponential Backoff and Geometric Retransmission systems provide the highest and the lowest robustness performance, respectively.

With an arbitrary cutoff phase 1<*K*<∞, the maximum stable throughput $\hat{\lambda}_{\max\_S\_q_m^{Exp}}^{K-Exp}$ can be derived according to Eqs. (50) and (82) in Part I. Since it does not have an explicit expression, we present the numerical results in Fig. 7.

As we can see from Fig. 7, the maximum stable throughput $\hat{\lambda}_{\max\_S\_q_m^{Exp}}^{K-Exp}$ sharply increases with the cutoff phase *K* and quickly approaches $e^{-1}$. When *K* is small, the system performance can be significantly improved even with a slight increment of *K*.

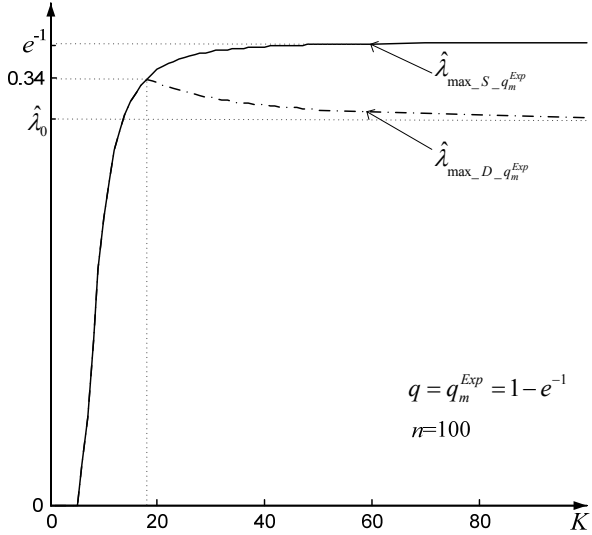

Fig. 7. The maximum stable throughput w/o delay constraint under different values of cutoff phase $K$ with $q=1-e^{-1}$ and $n=100$.

With a high traffic input rate, however, the queueing performance may start deteriorating as the cutoff phase $K$ increases. Suppose that the system operates at the desired stable point $p_L$ with $q=q_m^{Exp}$. We can see from (32) that when the traffic input rate $\hat{\lambda}$ exceeds $\hat{\lambda}_0$, we have

$$\sqrt{1-p_L} > q = 1-e^{-1}. \quad (38)$$

According to (5), the second moment of service time will exponentially grow with the cutoff phase $K$ and finally become unbounded as $K=\infty$.

Let us set the delay constraint to be $G''_{X_0}(1) < 1000$. The corresponding maximum stable throughput with the delay constraint, $\hat{\lambda}_{\max\_D\_q_m^{Exp}}^{K-Exp}$, under different values of $K$ is plotted in Fig. 7. It can be clearly seen that despite the boost in the maximum stable throughput $\hat{\lambda}_{\max\_S\_q_m^{Exp}}^{K-Exp}$, a further increase of cutoff phase $K$ may instead jeopardize the system performance. The highest maximum stable throughput with the delay constraint $\hat{\lambda}_{\max\_D\_q_m^{Exp}}^{K-Exp}$ is achieved at around 0.34 when $K=19$. Note that with a more stringent delay constraint, $\hat{\lambda}_{\max\_D\_q_m^{Exp}}^{K-Exp}$ may be even lower.

The above analysis is verified by simulation results presented in Table I. As we can see, the delay performance is not sensitive to the cutoff phase $K$ when the traffic rate is smaller than $\hat{\lambda}_0 \approx 0.3$. This is because the HOL packets are seldom pushed into deep phases when the traffic input rate is low. The maximum number of phase $K$ will not affect the delay performance as long as it is large enough to ensure a stable throughput.

With a high traffic rate, i.e., $\hat{\lambda} > \hat{\lambda}_0$, however, the mean queueing delay will rapidly increase with the cutoff phase $K$, and finally become infinite when $K=\infty$. As our analysis has revealed, although a larger cutoff phase $K$ leads to a higher maximum stable throughput, an excessively large $K$ will impair the queueing performance.

Table I. MEAN QUEUEING DELAY WITH *K*-EXPONENTIAL BACKOFF. $n=100$ and $q=1-e^{-1}$.

| $\hat{\lambda}$ | $K=10$ | $K=12$ | $K=20$ |
|---|---|---|---|
| 0.05 | 1.2 | 1.2 | 1.2 |
| 0.1 | 1.5 | 1.5 | 1.5 |
| 0.15 | 2.1 | 2.1 | 2.1 |
| 0.2 | 3.4 | 3.4 | 3.4 |
| 0.25 | 7.3 | 7.8 | 8.4 |
| 0.3 | ∞ | 39 | 185 |
| 0.35 | ∞ | ∞ | 5,562 |

### III. WINDOW-BASED *K*-EXPONENTIAL BACKOFF

The backoff model studied in Section II is slightly different from the contention-window-based backoff mechanism adopted in practical systems (such as Ethernet and IEEE 802.11 networks), where a HOL packet in phase $i$ selects a random value from its corresponding contention window. Instead of attempting with probability of $q^i$ at each time slot, the HOL packet does not retransmit until it counts down to zero. As mentioned in [8], many analytical studies focus on the retransmission-probability-based backoff model because of its memoryless nature [2-7, 21-24]. The performance analysis on practical systems, however, is based on the contention window mechanism [17-20].

In this section, we will extend the delay analysis to the window-based *K*-Exponential Backoff, and demonstrate that both backoff models achieve similar delay performance in most cases.

#### A. Window-based Backoff Model

A semi-Markov chain can be established to describe the state transition process of HOL packets in the window-based backoff model. As shown in Fig. 8, a HOL packet in phase $i$ will move to the next phase $i+1$ if it is involved in a collision, or jump back to phase 0 if it is successfully transmitted, $i=0,1,\ldots, K-1$. At each phase $i$, the HOL packet will randomly select a value from $\{0, \ldots, W_i-1\}$, where $W_i$ is the contention window size of phase $i$. It then counts down at each time slot and retransmits when the counter is zero.

The sojourn time at any phase $i=0,1,\ldots, K-1$, $Y_i$, is a uniformly distributed random variable with state space $\{1,\ldots, W_i\}$. At phase $K$, the HOL packet will stay until it is successfully transmitted. The sojourn time at phase $K$, $Y_K$, is therefore a renewal process and the inter-renewal interval follows the uniform distribution with state space $\{1,\ldots, W_K\}$.

Note that the probability-based backoff model discussed in Section II can be also regarded as a special case of the above semi-Markov chain. The sojourn time at each phase $i$, $Y_i$, is a geometric distributed random variable with parameter $q^i$ when $i=0,1,\ldots, K-1$ and $pq^K$ when $i=K$.

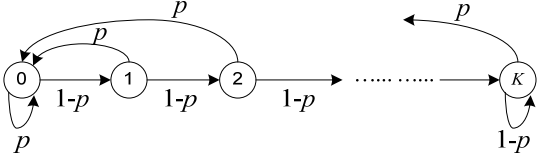

Fig. 8. Embedded Markov chain of the state transition process of HOL packets.

The limiting probabilities of the embedded Markov chain can be easily obtained as

$$f_i = p(1-p)^i, i=0,1,\ldots,K-1 \text{ and } f_K = (1-p)^K, \quad (39)$$

and the mean holding time at phase $i$ is

$$\tau_i = \frac{W_i+1}{2}, i=0,1,\ldots,K. \quad (40)$$

The phase distribution of HOL packets is then given by

$$\tilde{f}_i = \frac{f_i \cdot \tau_i}{\sum_{j=0}^{K} f_j \cdot \tau_j}, i=0,1,\ldots,K. \quad (41)$$

It can be seen from (39-41) that if the contention window size is set to be

$$W_i = 2/q^i - 1, i=0,1,\ldots,K, \quad (42)$$

the phase distribution can be obtained as

$$\tilde{f}_0 = 1 / \left( \frac{q}{p+q-1} - \left( \frac{q}{p+q-1} - \frac{1}{p} \right) \cdot \left( \frac{1-p}{q} \right)^K \right), \quad (43)$$

$$\tilde{f}_i = \tilde{f}_0 \left( \frac{1-p}{q} \right)^i, i=1,\ldots,K-1, \text{ and} \quad (44)$$

$$\tilde{f}_K = \tilde{f}_0 \left( \frac{1-p}{q} \right)^K / p, \quad (45)$$

which is consistent with Eqs. (3-5) in Part I derived based on the probability-based backoff model. In fact, the phase distribution of HOL packets is determined by the first moment of the sojourn time at each phase. Although different backoff models have distinct distributions of sojourn time $Y_i$, they follow the same phase distribution as long as the mean sojourn time $E[Y_i]$ is equal in both cases.

The stability analysis of probability-based $K$-Exponential Backoff has been presented in Part I of the paper series. The counterparts of Theorems 1 and 2 in the window-based case are shown in Appendix IV for demonstration. It can be easily checked that all the analytical results presented in Part I hold true for window-based $K$-Exponential Backoff, if the contention window size is set as (42). Both backoff models share the same stable region if they have the same mean sojourn time at each phase.

Note that the contention window size $W_i$ should be selected as an integer in practical systems, i.e.,

$$W_i = \lceil 2/q^i \rceil - 1, i=0,1,\ldots,K. \quad (46)$$

Here we ignore the difference of (42) and (46) for the sake of discussion. The simulation results will show that this approximation does not affect the conclusion.

*B. Delay Comparison of Window-based Backoff Model and Probability-based Backoff Model*

Let $X_i$ be the time spent from the beginning of phase $i$ until service completes. According to Fig. 8, the probability generating function of $X_i$ can be written as

$$\begin{cases} G_{X_i}(z) = pG_{Y_i}(z) + (1-p)G_{Y_i}(z)G_{X_{i+1}}(z), i=0,1,\ldots,K-1 \\ G_{X_K}(z) = G_{Y_K}(z). \end{cases} \quad (47)$$

where $G_{Y_i}(z)$ is the probability generating function of the sojourn time $Y_i$. It can be obtained from (47) that

$$G'_{X_0}(1) = \sum_{i=0}^{K} (1-p)^i G'_{Y_i}(1) \quad (48)$$

and

$$G''_{X_0}(1) = \sum_{i=0}^{K} (1-p)^i \left( G''_{Y_i}(1) + 2G'_{Y_i}(1) \sum_{j=1}^{K-i} (1-p)^j G'_{Y_{i+j}}(1) \right). \quad (49)$$

The mean queueing delay $E[T]$ can be then obtained by substituting (48-49) into (8).

The mean queueing delay is dependent on both the first and second moments of the sojourn time $Y_i$. The probability generating function of the sojourn time $Y_i$ in the window-based backoff model is given by

$$G_{Y_i}(z)_{win} = \frac{1}{W_i} \cdot \frac{z - z^{W_i+1}}{1-z}, i=0,1,\ldots,K-1 \quad (50)$$

and

$$G_{Y_K}(z)_{win} = \frac{p \cdot \frac{z - z^{W_K+1}}{1-z}}{1-(1-p)\frac{z-z^{W_K+1}}{1-z}}. \quad (51)$$

The first and second moments of $Y_i$ can be then obtained as

$$\begin{cases} G'_{Y_i}(1)_{win} = \frac{W_i+1}{2}, i=0,1,\ldots,K-1 \\ G'_{Y_K}(1)_{win} = \frac{W_K+1}{2p}. \end{cases} \quad (52)$$

and

$$\begin{cases} G''_{Y_i}(1)_{win} = \frac{(W_i+1)(W_i-1)}{3}, i=0,1,\ldots,K-1 \\ G''_{Y_K}(1)_{win} = \frac{(W_K+1)(W_K-1)}{3p} + \frac{(1-p)(W_K+1)^2}{2p^2}. \end{cases} \quad (53)$$

When the contention window size $W_i$ is set according to (42), it can be obtained from (53) that the second moment of $Y_i$, $i=0,1,\ldots,K-1$, in the window-based backoff model is given by

$$G''_{Y_i}(1)_{win} = \frac{4}{3} \cdot \frac{1-q^i}{q^{2i}} < \frac{2(1-q^i)}{q^{2i}} = G''_{Y_i}(1)_{prob} \quad (54)$$

which is smaller than that in the probability-based model. Similarly, we have

$$G''_{Y_K}(1)_{win} = \frac{2}{(pq^K)^2}\left(1 - \frac{p}{3} - \frac{2pq^K}{3}\right) < \frac{2(1-pq^K)}{(pq^K)^2} = G''_{Y_K}(1)_{prob} \quad (55)$$

Therefore, lower queueing delay can be expected in the window-based case. In the following, we will take the examples of Geometric Retransmission ($K=1$) and Exponential Backoff ($K=\infty$) to demonstrate the above results.

*1) Window-based Geometric Retransmission*

The first and second moments of service time $X$ of window-based Geometric Retransmission can be obtained by combining (48-49), (52-53), (42) and $K=1$:

$$G'_{X_0}(1)_{win} = 1 + \frac{1-p}{pq} \quad (56)$$

and

$$G''_{X_0}(1)_{win} = \frac{2(1-p)}{(pq)^2}\left(1 - \frac{p}{3} + \frac{pq}{3}\right). \quad (57)$$

Note that the window-based Geometric Retransmission has the same stable region as the probability-based one. With $q = q_m^{Geo} = 1/n$, or, equivalently,

$$W_0 = 1 \text{ and } W_1 = 2n-1, \quad (58)$$

the corresponding mean access delay and mean queueing delay are given by:

$$\mathrm{E}[X]^{Geo}_{W_1=2n-1} = 1 + n \cdot \frac{1-p_L}{p_L}, \quad (59)$$

$$\mathrm{E}[T]^{Geo}_{W_1=2n-1} = 1 + n\left(\frac{1}{p_L - \frac{\hat{\lambda}(1-p_L)}{1-\hat{\lambda}/n}}\left(1 - \frac{\hat{\lambda}(1-p_L)}{3(1-\hat{\lambda}/n)}\right) - 1\right). \quad (60)$$

It can be seen from (59-60) and (20-21) that despite the same access delay, lower queueing delay will be incurred in the window-based backoff model, and the delay gap will be enlarged as the traffic input rate $\hat{\lambda}$ or the number of nodes $n$ increases. This can be clearly observed in Fig. 9. Both backoff models have similar delay performance when the traffic input rate $\hat{\lambda}$ is low. The delay gap becomes noticeable with a large number of nodes $n$ and a high input rate $\hat{\lambda}$. Nevertheless, the difference is still quite small compared to the mean queueing delay $\mathrm{E}[T]$ incurred.

*2) Window-based Exponential Backoff*

The first and second moments of service time $X$ of window-based Exponential Backoff can be obtained by combining (48-49), (52-53), (42) and $K=\infty$:

$$G'_{X_0}(1)_{win} = 1 + \frac{1-p}{q-(1-p)} \quad (61)$$

and

$$G''_{X_0}(1)_{win} = \frac{2(1-p)q}{[q-(1-p)]\cdot[q^2-(1-p)]} \cdot \frac{2+q}{3} \quad (62)$$

if $q > \sqrt{1-p}$. With $q = q_m^{Exp} = 1-e^{-1}$, or, equivalently,

$$W_i = 2/(1-e^{-1})^i - 1, \ i=0, 1,\ldots, \quad (63)$$

and $\hat{\lambda} < \hat{\lambda}_0 \approx 0.3$, the mean access delay and mean queueing delay of window-based Exponential Backoff are given by

$$\mathrm{E}[X]^{Exp}_{W_i=2/(1-e^{-1})^i-1} = \frac{1-e^{-1}}{p_L - e^{-1}} \quad (64)$$

$$\mathrm{E}[T]^{Exp}_{W_i=2/(1-e^{-1})^i-1} = \frac{1-e^{-1}}{p_L - e^{-1}} + (1 - e^{-1}/3)$$

$$\cdot \frac{\hat{\lambda}(1-p_L)(1-e^{-1})}{(n(p_L - e^{-1}) - \hat{\lambda}(1-e^{-1}))(p_L - 2e^{-1} + e^{-2})} \quad (65)$$

As shown in Fig. 10, the delay gap between two backoff models is negligible with a traffic input rate $\hat{\lambda} < \hat{\lambda}_0 \approx 0.3$. In contrast to Geometric Retransmission, the gap does not increase with the number of nodes $n$ in the Exponential Backoff case, because the delay is not sensitive to the change of network population in both models.

When the traffic input rate $\hat{\lambda}$ is close to $\hat{\lambda}_0 \approx 0.3$, the delay performance rapidly deteriorates and the system becomes quasi-stable as $\hat{\lambda}$ exceeds 0.3. The capture phenomenon will occur regardless of which backoff model is adopted.

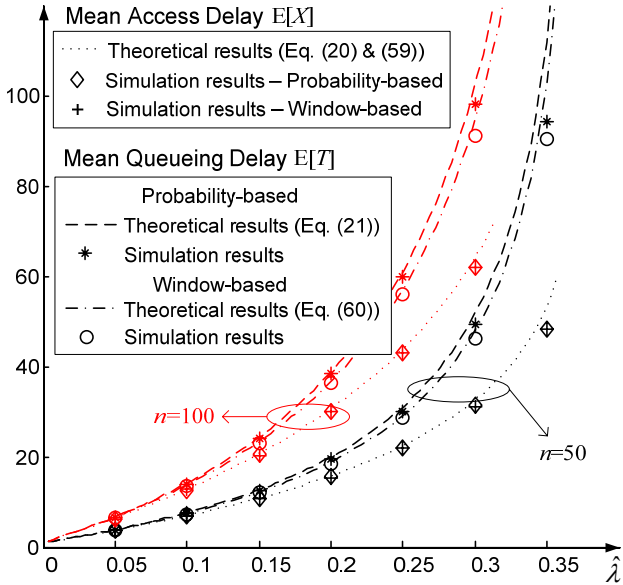

Fig. 9. Mean queueing delay and mean access delay versus aggregate input rate $\hat{\lambda}$ with Geometric Retransmission. $q=1/n$ in the retransmission-probability-based backoff model. $W_0=1$ and $W_1=2n-1$ in the contention-window-based backoff model.

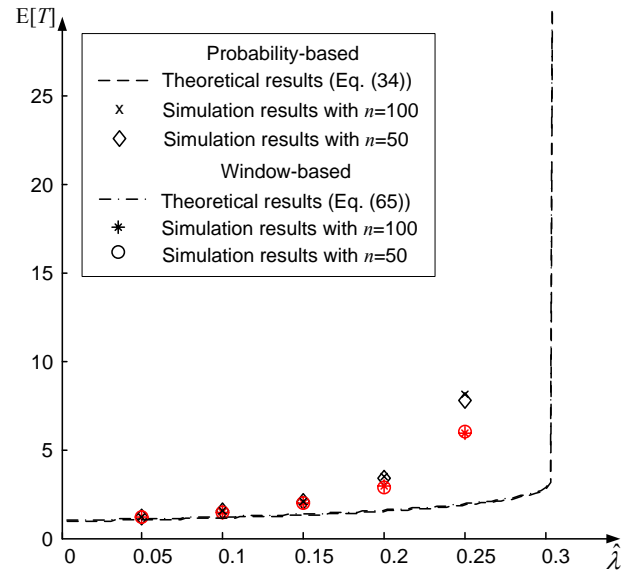

Fig. 10. Mean queueing delay versus aggregate input rate $\hat{\lambda}$ with Exponential Backoff. $q=1-e^{-1}$ in the retransmission-probability-based backoff model. $W_i = \lceil 2/(1-e^{-1})^i \rceil - 1$, $i=0,1,\ldots$ in the contention-window-based backoff model.

It can be clearly seen from the above analysis that similar queueing performance is achieved in both backoff models. In fact, it has been shown in (54-55) that the difference of the second moment of sojourn time at phase $i$ in the probability-based and the window-based models, i.e., $G''_{Y_i}(1)_{prob} - G''_{Y_i}(1)_{win}$, becomes significant only when the phase number $i$ is large. With Geometric Retransmission, the maximum number of phases is $K=1$. In the case of Exponential Backoff, despite an infinite cutoff phase $K$, the HOL packets usually stay at small phases when the traffic input rate $\hat{\lambda}$ is low. As a result, the delay gap between the two backoff models remains quite small all the while.

With $K$-Exponential backoff ($1<K<\infty$), similar queueing delay performance should be achieved in both backoff models when the cutoff phase $K$ is small. With a large $K$ and a high traffic input rate $\hat{\lambda}$, however, some HOL packets will be pushed into deep phases and the difference of the second moment of sojourn time of the two backoff models $G''_{Y_i}(1)_{prob} - G''_{Y_i}(1)_{win}$ will grow exponentially with the phase number $i$. Much lower queueing delay is therefore expected to be incurred by the window-based backoff model.

This can be clearly observed in Table II. Both the window-based and probability-based models share the same stable regions, and achieve almost the same queueing delay performance when the traffic input rate $\hat{\lambda}$ is low. With $\hat{\lambda}>0.3$, the delay gap becomes significant especially when the cutoff phase $K$ is large.

Table II. MEAN QUEUEING DELAY WITH CONTENTION-WINDOW-BASED AND RETRANSMISSION-PROBABILITY-BASED $K$-EXPONENTIAL BACKOFF. $n=100$, $q=1$-$e^{-1}$ and $W_i = \lceil 2/(1-e^{-1})^i \rceil - 1$, $i=0,1,...,K$.

| $\hat{\lambda}$ | $K=10$ | | $K=12$ | | $K=20$ | |
|---|---|---|---|---|---|---|
| | $W_i$ | $q^i$ | $W_i$ | $q^i$ | $W_i$ | $q^i$ |
| 0.05 | 1.2 | 1.2 | 1.2 | 1.2 | 1.2 | 1.2 |
| 0.1 | 1.5 | 1.5 | 1.5 | 1.5 | 1.5 | 1.5 |
| 0.15 | 2.0 | 2.1 | 2.0 | 2.1 | 2.0 | 2.1 |
| 0.2 | 2.9 | 3.4 | 2.9 | 3.4 | 2.9 | 3.4 |
| 0.25 | 5.7 | 7.3 | 5.9 | 7.8 | 6.2 | 8.4 |
| 0.3 | ∞ | ∞ | 25.6 | 39 | 75.5 | 185 |
| 0.35 | ∞ | ∞ | ∞ | ∞ | 4,520 | 5,562 |

## IV. CONCLUSIONS

The delay analysis of buffered Aloha networks with $K$-Exponential Backoff was presented in this paper. Geometric Retransmission ($K=1$) and Exponential Backoff ($K=\infty$) are demonstrated as exemplary cases, and the performance comparison shows that although both of them can achieve a stable throughput of $0<\hat{\lambda}\leq e^{-1}$ with the retransmission factor $q$ properly selected from the corresponding stable region, the delay performance is quite different. With Geometric Retransmission, finite mean queueing delay can be achieved within the whole traffic range $0<\hat{\lambda}\leq e^{-1}$, and both the mean access delay and the mean queueing delay linearly increase with the number of nodes $n$. The delay performance of Exponential Backoff is not sensitive to the change of network population; nevertheless, it is severely penalized when traffic is heavy. With the aggregate input rate $\hat{\lambda}>0.3$, the network will shift to the undesired stable point, and the variance of service time will grow unboundedly as time elapses, leading to infinite mean queueing delay. The cutoff phase $K$ can be carefully tuned to achieve a good tradeoff between queueing performance and system robustness in a general $K$-Exponential backoff network.

The delay analysis was further extended to the contention-window-based backoff case. The performance comparison indicates that the retransmission-probability-based backoff model can serve as a good analytical model of the practical contention window mechanism, as it follows the same stable region and provides a tight delay upperbound.

## APPENDIX I. POLLACZEK-KINTCHINE FORMULA OF GEO/G/1

Let $A$ represent the number of arrivals during service time $X$. The probability generating function of $A$ is then given by

$$M_A(z) = \sum_{i=0}^{\infty} z^i \sum_{k=1}^{\infty} \Pr\{A=i \mid X=k\} \Pr\{X=k\}$$
$$= \sum_{i=0}^{\infty} z^i \sum_{k=1}^{\infty} \binom{k}{i} \lambda^i (1-\lambda)^{k-i} \Pr\{X=k\}$$
$$= M_X(1-\lambda+\lambda z) \qquad (66)$$

where $\lambda$ is the input rate.

Let $C_n$ be the number of customers left behind by the departure of the $n$-th customer, and $A_n$ be the number of customers arriving during the service of the $n$-th customer. The length of the buffer queue is given by the Lindley equation:

$$C_{n+1} = C_n - I(C_n) + A_{n+1} \qquad (67)$$

where

$$I(C_n) = \begin{cases} 0 & \text{if } C_n = 0 \\ 1 & \text{if } C_n > 0 \end{cases} \qquad (68)$$

is the indicator function. Let $C = \lim_{n\to\infty} C_n$. We have

$$M_C(z) = E[z^{C-I(C)+A}] = \left[\left(1-\frac{1}{z}\right)p_0 + \frac{1}{z}M_C(z)\right] \cdot M_A(z)$$
$$\Rightarrow M_C(z) = \frac{p_0(z-1)M_A(z)}{z-M_A(z)}. \qquad (69)$$

where $p_0=1-\rho$ and $\rho$ is the offered load.

Let $T$ denote the queueing delay of each packet. We have

$$\Pr\{C=k\} = \sum_{n=1}^{\infty} \Pr\{A=k \mid T=n\} \Pr\{T=n\}. \qquad (70)$$

The probability generating function of $C$ can then be written as

$$M_C(z) = \sum_{k=0}^{\infty} z^k \sum_{n=1}^{\infty} \binom{n}{k} \lambda^k (1-\lambda)^{n-k} \Pr\{T=n\}$$
$$= \sum_{n=1}^{\infty} \Pr\{T=n\} \sum_{k=0}^{n} \binom{n}{k} \lambda^k (1-\lambda)^{n-k} z^k$$
$$= \sum_{n=1}^{\infty} \Pr\{T=n\}(1-\lambda+\lambda z)^n = M_T(1-\lambda+\lambda z). \qquad (71)$$

Combining (66), (69) and (71), we have

$$M_T(z) = \frac{p_0(z-1)M_X(z)}{(z-1)+\lambda[1-M_X(z)]}. \quad (72)$$

From (72) finally we can obtain the mean queueing delay of a Geo/G/1 queue as

$$E[T] = E[X] + \frac{\lambda\{\text{var}[X]+E[X]^2-E[X]\}}{2(1-\lambda E[X])}. \quad (73)$$

## APPENDIX II.  $q < \sqrt{1-p_A^{Exp}}$ FOR LARGE $n$

It has been derived in Appendix II of Part I that the undesired stable point of Exponential Backoff is given by

$$p_A^{Exp} = \frac{n(1-q)/q}{W_0((n/q-n)\exp(n/q))} \approx \frac{n(1-q)}{n+q\ln(1-q)}. \quad (74)$$

For given $0<q<1$, as long as the number of nodes $n$ satisfies

$$n > -(1+q)\ln(1-q), \quad (75)$$

we have

$$\sqrt{1-p_A^{Exp}} = \sqrt{q \cdot \frac{n+\ln(1-q)}{n+q\ln(1-q)}} > q. \quad (76)$$

## APPENDIX III. LOWER BOUND OF INSTANTANEOUS ATTEMPT RATE OF EXPONENTIAL BACKOFF WITH A LARGE NUMBER OF NODES *n*

In an *n*-node buffered Exponential Backoff network, suppose that there are totally $n_b$ backlogged HOL packets at time slot $t$, with $n_i$ packets in phase $i$, $i=1, 2,\ldots$. The attempt rate $G_t$ is given by

$$G_t = (n-n_b)\lambda + \sum_{i=1}^{\infty} n_i q^i \geq \sum_{i=1}^{\infty} n_i q^i, \quad (77)$$

where

$$\sum_{i=1}^{\infty} n_i = n_b. \quad (78)$$

Let $\phi_i$ denote the probability that there is a phase-*i* HOL packet in the node's buffer, $i=1, 2,\ldots$. We know from the Markov chain shown in Fig. 2 of Part I that

$$\phi_i = \rho f_i, \quad i=1, 2,\ldots \quad (79)$$

where $\rho$ is the offered load of each queue and $f_0, f_1, \ldots$ are the limiting probabilities, and we have

$$f_0 \geq f_1 \geq f_2 \geq \cdots \quad (80)$$

According to the law of large numbers, at any time slot $t$, the number of backlogged HOL packets in phase $i$, $n_i$, satisfies

$$\frac{n_i}{n} \to \phi_i \quad \text{for } n\to\infty, \quad (81)$$

and

$$\frac{n_b}{n} \to \rho(1-f_0) \quad \text{for } n\to\infty. \quad (82)$$

By combining (79-81), we can conclude that for large *n*, there is a high probability that

$$n_1 \geq n_2 \geq n_3 \geq \cdots \quad (83)$$

As a result, the attempt rate $G_t$ is lower bounded by

$$G_t \geq \sum_{i=1}^{\infty} n_i q^i \geq \sum_{i=1}^{n_b} q^i = \frac{q-q^{n_b}}{1-q} \to \frac{q}{1-q} \quad \text{for } n\to\infty. \quad (84)$$

With a large number of nodes *n*, if the retransmission factor $q$ satisfies (28), we can see from (84) that the attempt rate $G_t$ will exceed $-\ln p_S$ with high probability, and the probability of success $p$ will converge to the undesired stable point $p_A$.

## APPENDIX IV.  PROOFS OF THEOREMS 1 AND 2

We need the following lemma before presenting the proofs of Theorems 1 and 2.

**Lemma 1.** *For buffered Aloha with window-based K-Exponential Backoff, the probability that a phase-i HOL packet has a transmission request, i=0,1,…, K, is given by*

$$r_i = \frac{2}{W_i+1}. \quad (85)$$

Proof: In the window-based backoff model, each HOL packet in phase $i$ chooses a random value from $\{0, \ldots, W_i-1\}$, $i=1,\ldots, K$, with equal probability, and sends a transmission request when counting down to zero. Let $J_t^i$ denote the residual time at time slot $t$ of a phase-*i* HOL packet. The transition process of $\{J_t^i\}$ can be described by the Markov chain shown in Fig. 11.

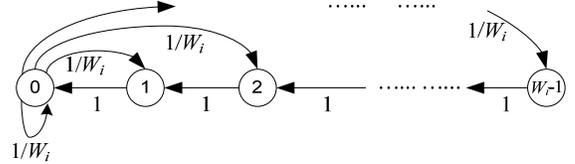

Fig. 11. State transition diagram of the residual time of a phase-*i* HOL packet.

The limiting probabilities of the above Markov chain can be obtained as

$$\pi_k = \frac{2}{1+W_i} \cdot \frac{W_i-k}{W_i}, \quad k=0, 1, \ldots, W_i-1. \quad (86)$$

The HOL packet will make a transmission request when the residual time becomes 0. As a result, the probability that a phase-*i* HOL packet has a transmission request, $r_i$, is given by

$$r_i = \pi_0 = \frac{2}{1+W_i}, \quad i=0, 1, \ldots, K. \quad \square$$

According to (42) and Lemma 1, the request probability of a phase-*i* HOL packet is given by

$$r_i = \frac{2}{W_i+1} = q^i, \quad i=0, 1, \ldots, K. \quad (87)$$

The counterparts of Theorems 1 and 2 in Part I can be then presented in the following.

**Theorem 1.** *For buffered Aloha with window-based K-exponential backoff ($1\leq K\leq\infty$), the probability of success p is given by*

$$p = \exp(-\hat{\lambda}/p), \quad (88)$$

*as the number of nodes $n\to\infty$, where $\hat{\lambda} = n\lambda$ is the aggregate input rate.*

Proof: Each node in the network must be in one of the following states:
State 1: idle;

State 2: busy with a fresh HOL;
State 3: phase $i$ and retransmitting, $i=1,2,\ldots,K$;
State 4: phase $i$ and not retransmitting, $i=1,2,\ldots,K$.

The probability of a HOL packet being in phase $i$ given that the node is busy is $\tilde{f}_i$, $i=0,1,\ldots,K$. Suppose that the offered load per queue is $\rho<1$. The probability of the above four states are given by:

1) Pr{node is in State 1}=$1-\rho$;
2) Pr{node is in State 2}=$\rho\tilde{f}_0$;
3) Pr{node is in State 3}=$\rho\tilde{f}_i r_i$, $i=1,2,\ldots K$;
4) Pr{node is in State 4}=$\rho\tilde{f}_i(1-r_i)$, $i=1,2,\ldots K$.

When a node successfully transmits a packet, its $n-1$ interfering nodes must be either in State 1 or State 4. The probability of success $p$ in steady-state conditions can then be written as

$$p = \left(\text{Pr\{node is in State 1\}} + \sum_{i=1}^{K}\text{Pr\{node is in State 4, phase } i\}\right)^{n-1}$$
$$= \left(1-\rho + \sum_{i=1}^{K}\rho\tilde{f}_i(1-r_i)\right)^{n-1} = \left(1-\rho(\tilde{f}_0+\sum_{i=1}^{K}\tilde{f}_i r_i)\right)^{n-1}. \quad (89)$$

Note that the offered load

$$\rho = \lambda/\tilde{f}_0 \quad (90)$$

as $r_0=1$. By combining (43-45), (87) and (89-90), we have

$$p = (1-\lambda/p)^{n-1} \to \exp(-\hat{\lambda}/p) \text{ for } n\to\infty. \quad \Box$$

**Theorem 2.** *For buffered Aloha with window-based K-Exponential Backoff ($1\leq K\leq\infty$), if $q \leq -\frac{1}{n}\ln p_S$, then at any time slot t, $G_t \leq -\ln p_S$.*

Proof: Suppose that there are totally $n_b$ backlogged HOL packets at time slot $t$, with $n_i$ packets in phase $i$, $i=1,\ldots,K$. The attempt rate $G_t$ is then given by

$$G_t = (n-n_b)\lambda + \sum_{i=1}^{K}n_i r_i. \quad (91)$$

According to (86), (91) can be written as

$$G_t = (n-n_b)\lambda + \sum_{i=1}^{K}n_i q^i \leq (n-n_b)\lambda + n_b q. \quad (92)$$

1) If retransmission factor $q\leq\lambda$, the attempt rate $G_t$ is bounded by

$$G_t \leq \hat{\lambda} < -\ln p_S. \quad (93)$$

2) If retransmission factor $q\geq\lambda$, the attempt rate $G_t$ is bounded by

$$G_t \leq nq \leq -\ln p_S. \quad (94)$$

Hence, the theorem is established by combining (93) and (94).$\Box$